\pdfoutput=1
\documentclass{JINST}
\usepackage{cite}
\newcommand*{\m}{\mathrm}

\title{On the Electric Breakdown in Liquid Argon at Centimeter Scale}

\author{M.~Auger, A.~Blatter, A.~Ereditato, D.~Goeldi\thanks{Corresponding author.}~, S.~Janos, I.~Kreslo, M.~Luethi, C.~Rudolf~von~Rohr, T.~Strauss\thanks{Now at FNAL, Batavia, IL.}~, and M.~S.~Weber\\
Laboratory for High Energy Physics\\
Albert Einstein Center for Fundamental Physics\\
Universit\"{a}t Bern, Switzerland\\
E-mail: \email{goeldi@protonmail.ch}}

\abstract{We present a study on the dependence of electric breakdown discharge properties on electrode geometry and the breakdown field in liquid argon near its boiling point.
The measurements were performed with a spherical cathode and a planar anode at distances ranging from 0.1~mm to 10.0~mm.
A detailed study of the time evolution of the breakdown volt-ampere characteristics was performed for the first time.
It revealed a slow streamer development phase in the discharge.
The results of a spectroscopic study of the visible light emission of the breakdowns complement the measurements.
The light emission from the initial phase of the discharge is attributed to electro-luminescence of liquid argon following a current of drifting electrons.
These results contribute to set benchmarks for breakdown-safe design of ionization detectors, such as Liquid Argon Time Projection Chambers (LAr~TPC).}


\begin{document}
\section{Introduction}

The practical implementation of large, long-drift distance time projection chambers based on liquid argon (LAr~TPCs), such as MicroBooNE~\cite{uboone}, ARGONTUBE~\cite{ARGONTUBE0,ARGONTUBE1,ARGONTUBE2} or the proposed DUNE~\cite{DUNE}, is driven by the effective dielectric strength of the cryogenic medium.
The high electrical potential values envisioned for such detectors (from 100~kV to 1~MV) require special care to avoid risks of electric breakdown in regions typically outside the drift volume.
Recently, we have reported in~\cite{Blatter:2014wua} the observation of breakdowns in liquid argon for electric field intensities as low as 40~kV/cm at the surface of mechanically polished stainless steel cathodes of 4~cm and 8~cm diameter.
The breakdowns occurred at cathode-anode distances of the order of 1~cm.
In a follow-up paper~\cite{latex}, we reported a method to increase the breakdown field strength by an order of magnitude.
In this paper, we finally present the results of a thorough study of the breakdown development in liquid argon based on its light emission in the visible range and on the measurement of its volt-ampere characteristics.

\section{Experimental setup\label{setup}}

\begin{figure}[htb]
\centering	
\includegraphics[width=0.265\linewidth]{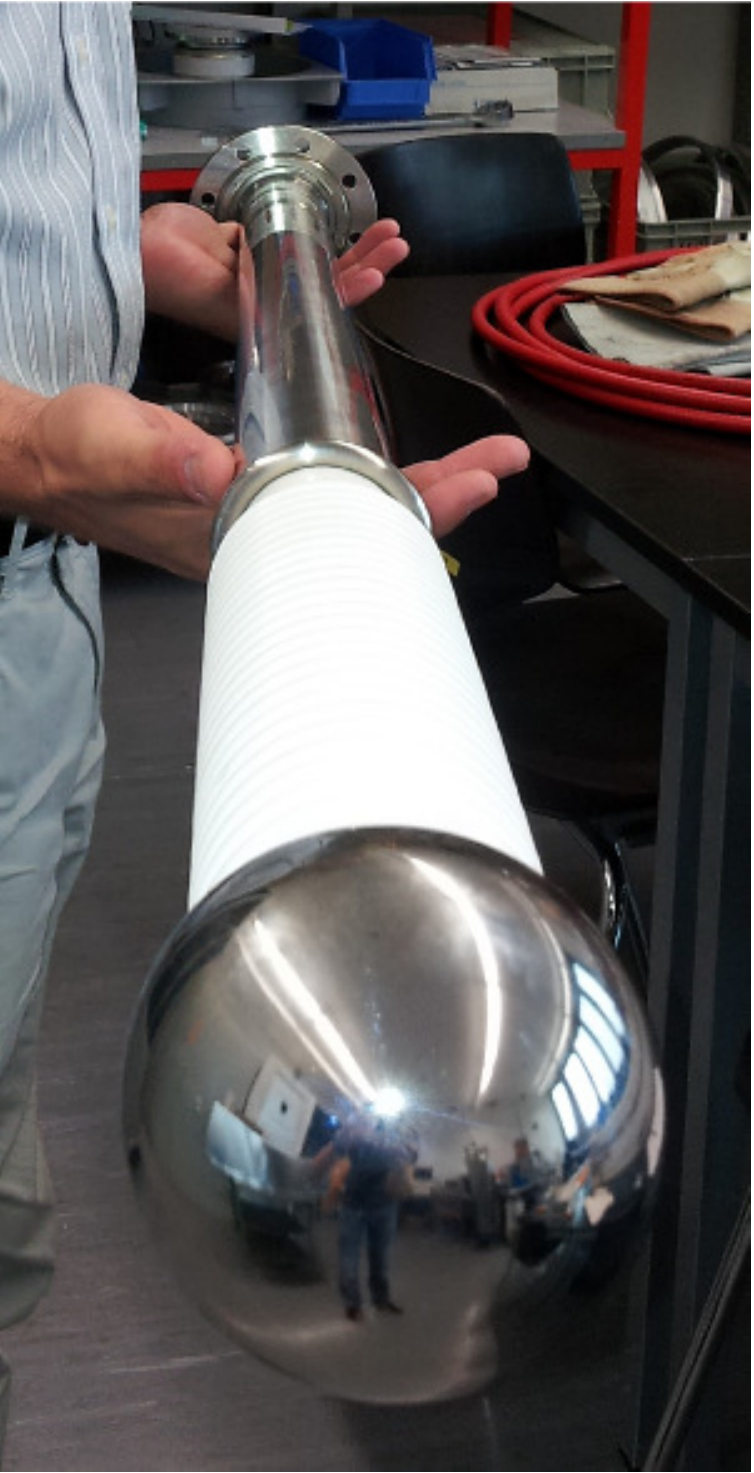}
\includegraphics[width=0.39\linewidth]{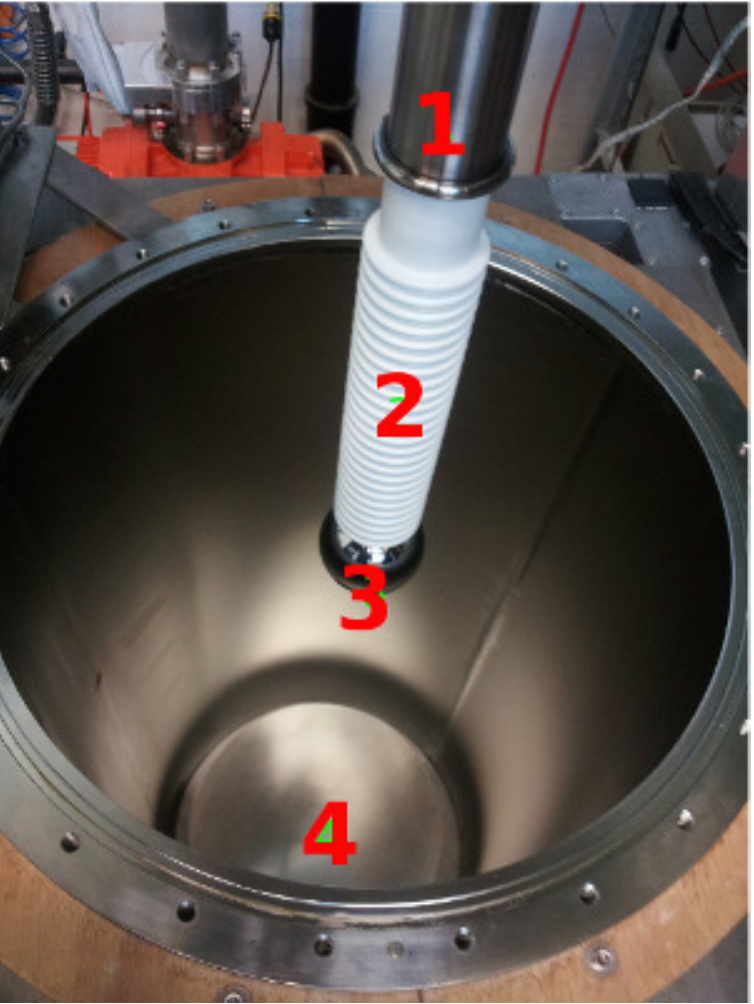}
\includegraphics[width=0.295\linewidth]{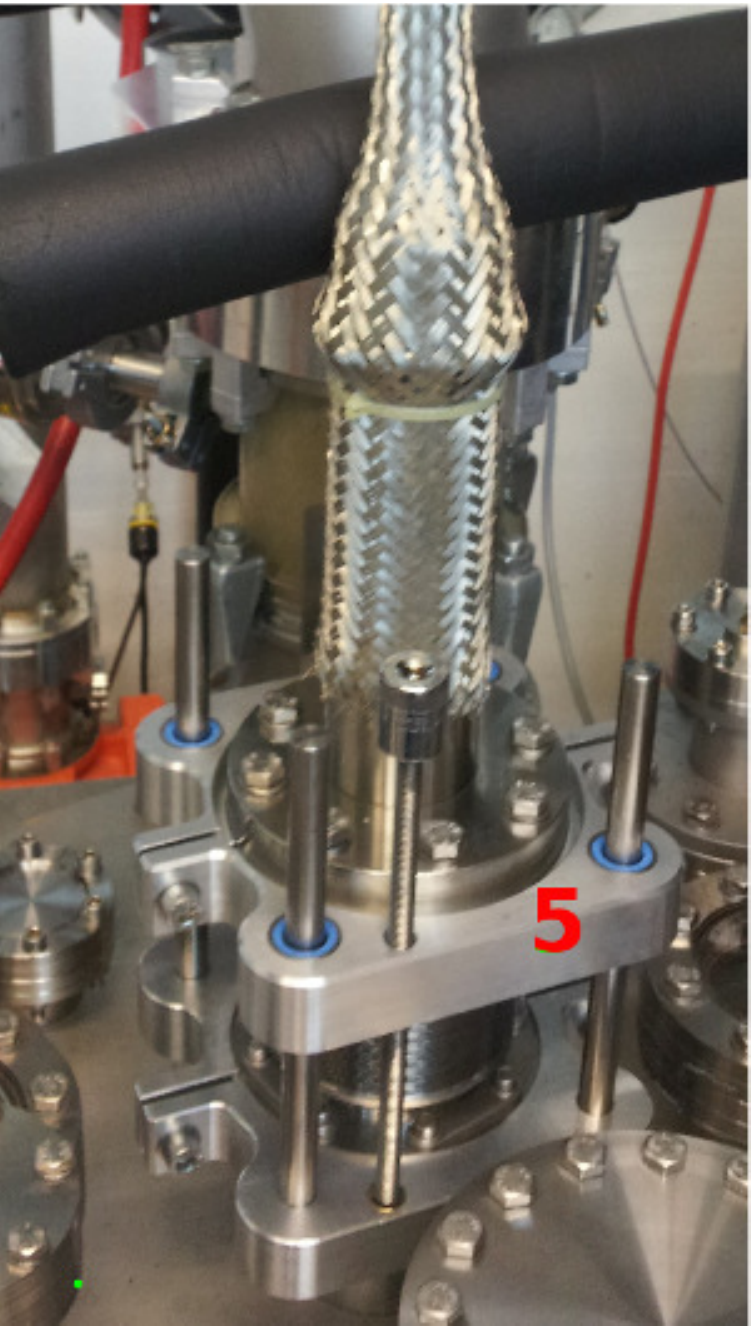}
\caption{Experimental setup. Left: high voltage feed-through with the spherical cathode. Middle: the feed-through before insertion into the cryostat. 1.~ground shield of the feed-through; 2.~ribbed PET dielectric; 3.~cathode sphere; 4.~anode plate sitting on a tripod on the grounded bottom of the cryostat; two of the tripod legs are insulated while the third one contains a 50~$\Omega$ shunt resistor. Right: linear translation unit used to set the cathode-anode gap width~(5).}
\label{fig:setup1}
\end{figure}

\begin{figure}[htb]
\centering	
\includegraphics[width=.99\linewidth]{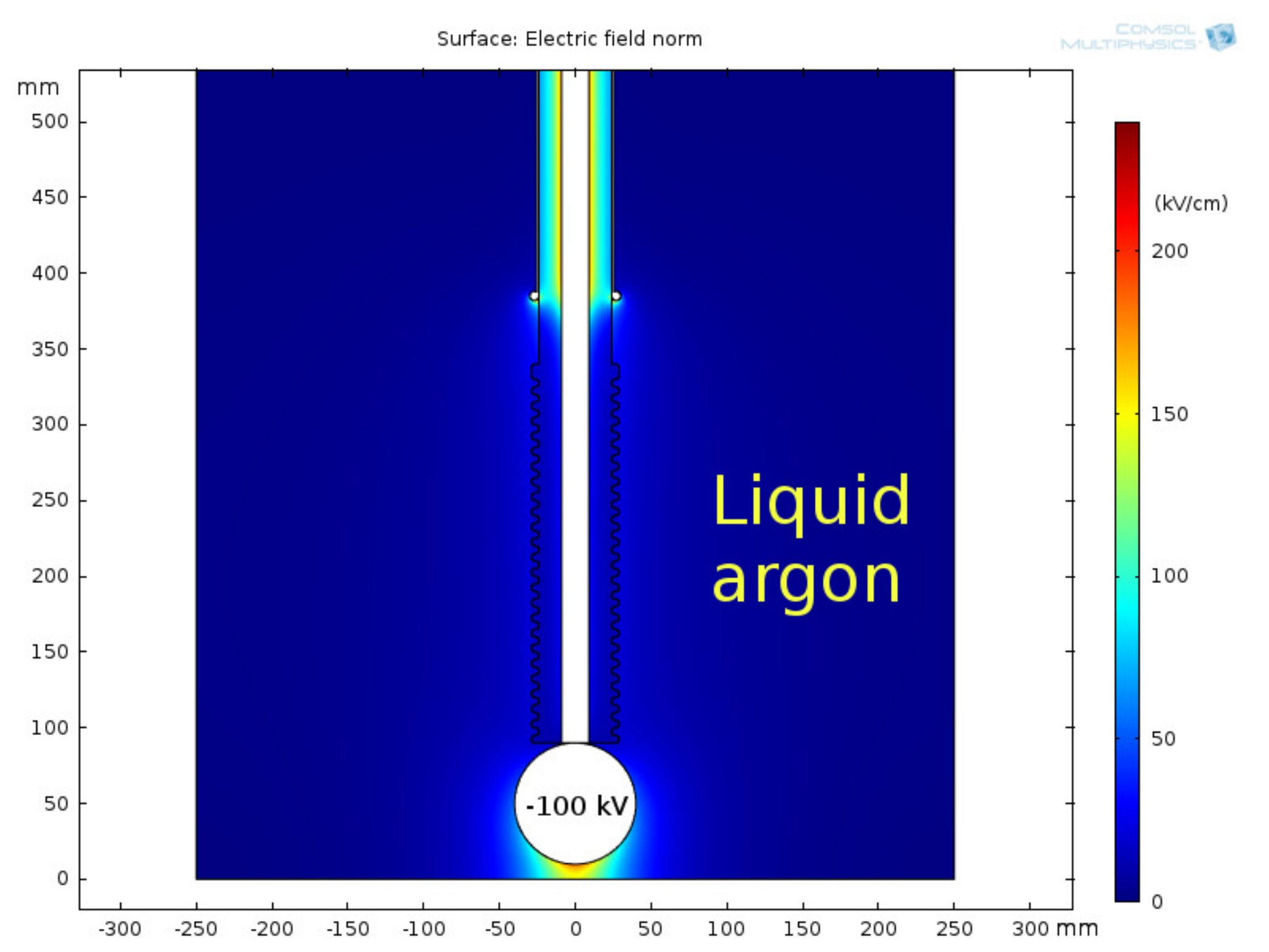}
\caption{Calculated electric field-amplitude map for the test setup with -100~kV at the cathode and cathode-anode distance of 1~cm.}
\label{fig:efield}
\end{figure}

The setup we used in this study is very similar to the one already described in~\cite{Blatter:2014wua} and is shown in Figure~\ref{fig:setup1}.
A spherical cathode and a plane anode electrode form the discharge gap.
Three different diameters of the cathode sphere were tested: 4~cm, 5~cm and 8~cm.
Two types of surface treatment were used in the cathode preparation, namely mechanical fine polishing and electro-polishing.
For the anode, mechanical fine polishing was used for all measurements.
The anode-cathode gap width can be set in the range of 0~mm to 100~mm with a precision of 0.3~mm.
An example of the field distribution in the setup is shown in Figure~\ref{fig:efield}.
The field map was calculated using the COMSOL FEM package.\footnote{\href{http://www.comsol.com}{http://www.comsol.com}}

The purity of the argon after filling was estimated with a small time projection chamber (according to the method described in~\cite{Badhrees:2010zz}) to be of the order of 1~ppb of oxygen-equivalent impurity concentration.
More details of the setup can be found in~\cite{Blatter:2014wua}.

\begin{figure}[htb]
\centering
\includegraphics[width=0.7\linewidth, angle=-90, origin=c]{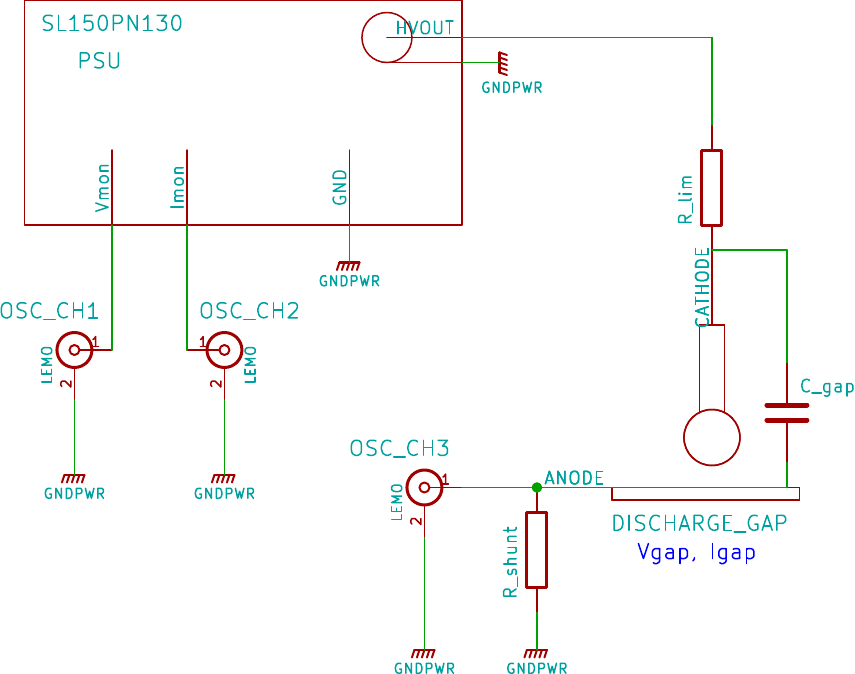}
\caption{Electric scheme of the experimental setup. The oscilloscope is connected to the control circuit of the high voltage power supply and to a shunt resistor on the ground return path. From the recorded voltages, the voltage and the current of the discharge can be derived.}
\label{fig:scheme}
\end{figure}

The control circuit of the \emph{Spellman SL150}\footnote{\href{http://www.spellmanhv.com}{http://www.spellmanhv.com}} power supply unit (PSU) outputs two low voltages proportional to the voltage and the current at the output, respectively.
These voltages are recorded with a \emph{Tektronix DPO 3054}\footnote{\href{http://www.tek.com}{http://www.tek.com}} digital oscilloscope.
The output polarity of the PSU can be switched by replacing the output HV multiplier module.
To measure the discharge current, a 50~$\Omega$ shunt resistor is placed between the anode plate and the vessel ground which is connected to the ground return of the PSU. The voltage drop across the shunt resistor is transmitted via a matched coaxial line to the oscilloscope.
The latter is controlled by a LabVIEW program.
The equivalent electric scheme of the setup is shown in Figure~\ref{fig:scheme}. 
The voltage at the $V_{\m{mon}}$ output of the power supply corresponds to the PSU output voltage divided by a factor $K_{\m{V}}$.
The voltage at $I_{\m{mon}}$ is related to the PSU output current.
However, according to the manufacturer of the PSU, an accurate reconstruction of the output current for frequencies above 100~Hz is not possible because of a filtering circuit in the current control loop.
Therefore, for the measurement of the discharge current only the voltage drop across the shunt resistor is used.

The oscilloscope is triggered by the channel connected to the shunt resistor.
The breakdown discharge current is composed of the output current of the PSU and the discharge current of the setup capacitance $C_{\m{gap}}$ as $I_{\m{gap}} = I_{\m{out}} + C_{\m{gap}}\frac{\m{d}V_{\m{gap}}}{\m{d}t}$.
To limit the PSU output current an additional resistor R$_{lim}$ is inserted into the HV output circuit.
The measured values for the circuit parameters are summarized in Table~\ref{table1}.
The knowledge of these parameters allows to calculate the voltage across the gap during breakdown. 

\begin{table}[htb]
\centering
\caption{Summary of the measured parameters of the test circuit.}
\label{table1}
\begin{tabular}{|l|l|}
\hline
$K_{\m{V}}$ & 		42.3$\cdot$10$^{-6}$ \\
\hline
$R_{\m{shunt}}$ &	50~$\Omega$ \\
\hline
$R_{\m{lim}}$ & 	200~M$\Omega$ \\
\hline
$C_{\m{gap}}$ & 	370 pF \\
\hline
\end{tabular}
\end{table}

In addition, the setup is equipped with an \emph{AOS Technologies S-PRI\emph{Plus}}\footnote{\href{http://www.aostechnologies.com}{http://www.aostechnologies.com}} high-speed camera to observe the development of the discharge. The camera is capable of recording $700 \times 400$ pixel RGB images at 1250~fps.
The camera comprises a frame ring buffer and is triggerable by an external TTL pulse.
This allows a synchronous recording of the visual appearance of the discharge and its volt-ampere characteristics.
The camera is triggered from the TRIG output of the oscilloscope.
The luminous part of the discharge is analysed for each frame of the recorded sequence.

The camera is mounted above a 5~cm diameter glass view port, located at the top flange of the cryostat and is looking downward.
To observe a discharge from the side, a glass mirror plate is installed at the edge of the cathode plane located 20~cm from the cathode, in such a way as to not perturb the electric field in the discharge gap. 

Finally, a custom built optical spectrometer is used to analyze the light emission of the discharges.
The spectrometer is connected to an optical fiber entering the cryostat with its other end attached to the anode plate.
The fiber is aligned such that its end directly faces the discharge gap resulting in a high angular acceptance.
As we show later, the discharge emission spectra allow to better understand the processes at different stages of the discharge.
This is due to the fact that the emission spectra of excited neutral atomic argon, singly-ionized and multiply-ionized atoms lay in different regions of the visible spectrum.

The possibility of creating gas bubbles near the discharge gap is inhibited by keeping the pressure in the inner vessel at 100~mbar above atmospheric pressure, plus an additional 100~mbar due to the hydro-static pressure.
The outer bath is opened to the atmosphere, thus keeping the inner vessel temperature constant and well below the boiling point.
No boiling was detected anywhere near the discharge gap region during the measurements.

\section{Experimental results}

\begin{figure}[htb]
\centering	
\includegraphics[width=.99\linewidth]{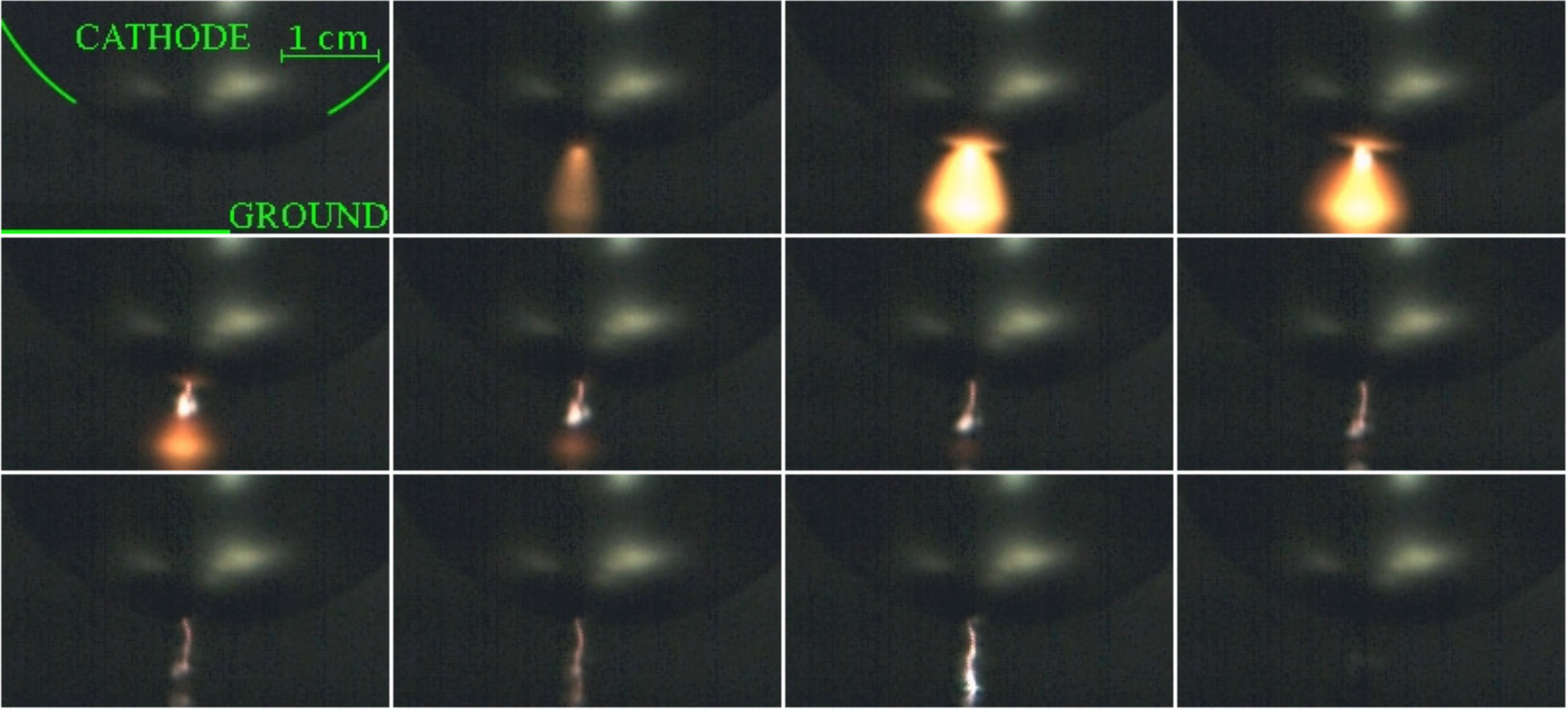}
\caption{Recorded camera image sequence for a breakdown from a 5~cm diameter cathode at -100.0~kV and 8.0~mm apart from the anode plate. The sequence is taken at 1250 fps, each frame takes 0.8~ms.}
\label{fig:images}
\end{figure}

In our earlier measurements~\cite{Blatter:2014wua}, we experienced sporadic discharges across the ribs of the dielectric of the high voltage feed-through.
In the present study, we managed to suppress these discharges completely by rising the level of the liquid argon by about 20~cm.
This improved the cooling of the feed-through and reduced bubble production near the bottom of the feed-through grounded shield, placed 60~cm below the liquid surface. 

The measurement campaign comprised 5~runs with a total of more than 5000 measured discharges, with varying sphere diameter, surface treatment and polarity.
A summary is shown in Table~\ref{table2}.
A typical recorded camera image sequence for breakdown from a 5~cm diameter cathode at -100.0~kV and 8.0~mm apart from the anode plate is shown in Figure~\ref{fig:images}.
The movie can be found as \emph{movie1.webm} in the ancillary files of this article.\footnote{\href{http://arxiv.org/src/1512.05968v2/anc/movie1.webm}{http://arxiv.org/src/1512.05968v2/anc/movie1.webm}}

\begin{table}[htb]
\centering\caption{Summary of the measurement runs.}
\label{table2}
\begin{tabular}{|l|l|l|l|l|l|l|}
\hline
Run &	$\oslash_{\m{Sphere}}$ &	Surface finish &	Sphere pol. &	Events &	$d_{\m{Gap}}$ &		$V_{\m{Breakdown}}$ \\
\hline
1   &	4~cm &						Mech.\ polished &	Cathode (-) &	1086 &		0.5~mm--8.0~mm &	3~kV--130~kV \\
\hline
2   &	5~cm &						Mech.\ polished &	Cathode (-) &	900 &		0.2~mm--12.0~mm &	2~kV--130~kV \\
\hline
3   &	8~cm &						Mech.\ polished &	Cathode (-) &	2434 &		0.1~mm--70.0~mm &	1~kV--130~kV \\
\hline
4   &	5~cm &						Mech.\ polished &	Anode (+) &		102 &		4.0~mm--5.0~mm &	5~kV--114~kV \\
\hline
5   &	5~cm &						Electro-polished &	Cathode (-) &	1141 &		0.1~mm--10.0~mm &	1~kV--130~kV \\
\hline
\end{tabular}
\end{table}

\begin{figure}[p]
\centering
\includegraphics[height=0.43\textheight]{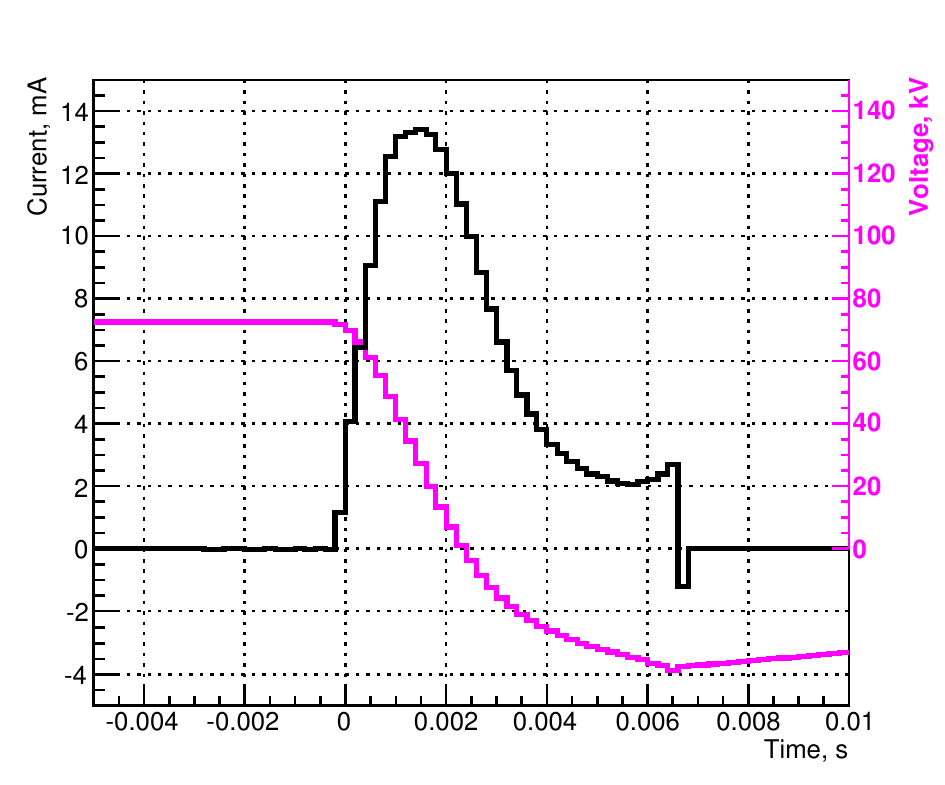}\\
\includegraphics[height=0.43\textheight]{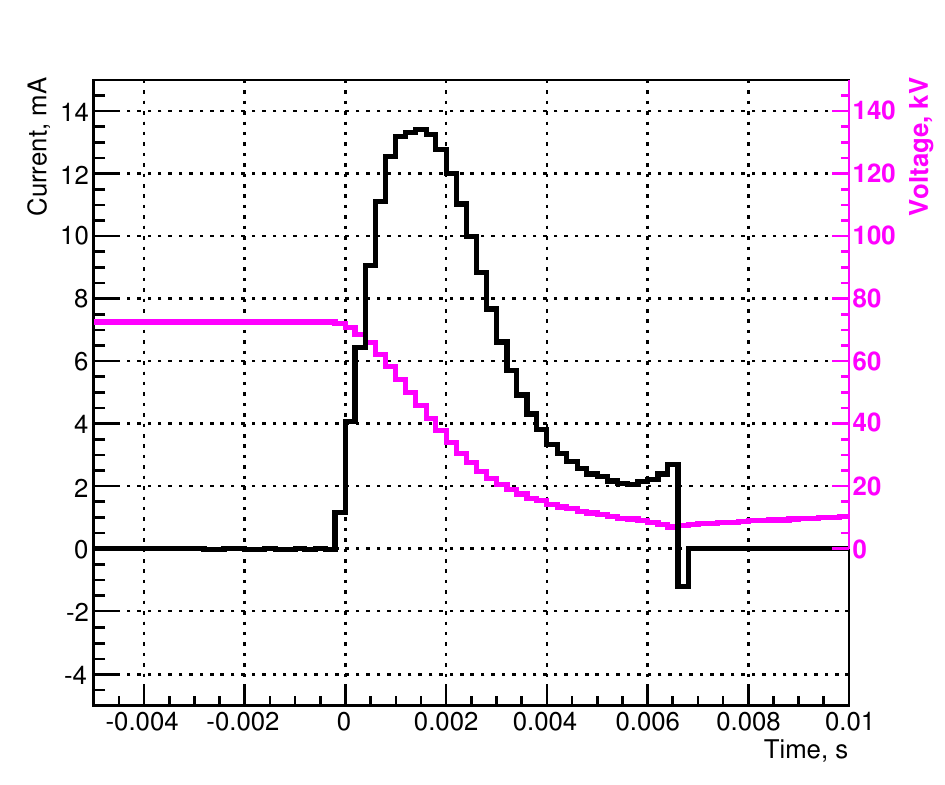}
\caption{Measured current through the gap (black) and voltage across the gap (magenta) for a typical breakdown at 6.0~mm distance between a 4~cm diameter cathode and the anode plate. The top plot shows the voltage obtained using the measured values of the protection resistor and the gap capacitance, while the bottom plot uses the tuned values.}
\label{fig:iv}
\end{figure}
\clearpage

In Figure~\ref{fig:iv}, we show the current and voltage features of a similar breakdown from a 4~cm diameter cathode at 6.0~mm from the anode plate.
The current was directly measured by observing the voltage drop across the shunt resistor, while the voltage was obtained by integrating the current taking into account the gap capacitance, the protection resistor and the output voltage of the power supply.
When using the measured values of Table~\ref{table1} for capacitance and resistance, this results in a negative voltage at the end of most discharges.
This is an unphysical result as can be seen in the top plot of Figure~\ref{fig:iv}.
This behavior may be attributed to poor knowledge of the effective values of the current-limiting resistor and the setup capacitance in the frequency domain of the discharge.
In order to better approximate these parameters, they are tuned in such a way that the minimum voltage for a maximum number of discharges approaches zero.
The best result was achieved by lowering the resistance by a factor of 1.7 and increasing the capacitance by the same factor.
Interestingly, this result leaves the RC characteristics of the system unchanged.
The bottom plot of Figure~\ref{fig:iv} shows the result obtained by using the tuned values for capacitance and resistance.

\begin{figure}[htb]
\centering	
\includegraphics[width=.75\linewidth]{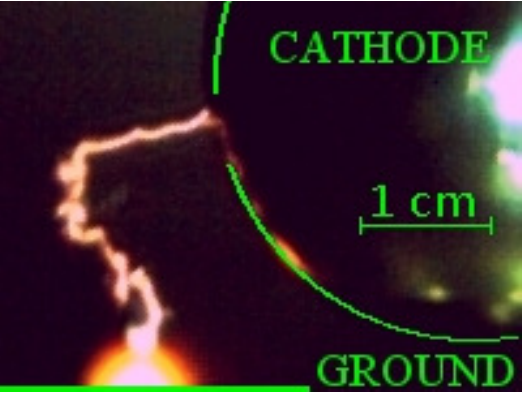}
\caption{An image of the streamer stage of the discharge, initiated at the side surface of a 5~cm cathode sphere at -121.5~kV and 8.0~mm apart from the anode plate. The cone of electrons emitted into liquid from the streamer tip towards the anode (lower edge of the image) produces bright orange luminescence in liquid argon.}
\label{fig:side}
\end{figure}

Most of the discharges are localized in the area of high field concentration between the tip of the sphere and the anode plane.
However, in rare cases, the discharge is initiated far from that region, sometimes at the side surface of the sphere.
An example of such a discharge is shown in Figure~\ref{fig:side} and the corresponding movie can be found as \emph{movie2.webm} in the ancillary files of this article.\footnote{\href{http://arxiv.org/src/1512.05968v2/anc/movie2.webm}{http://arxiv.org/src/1512.05968v2/anc/movie2.webm}}

\section{Interpretation of the results}

\begin{figure}[htb]
\centering
\includegraphics[width=.99\linewidth]{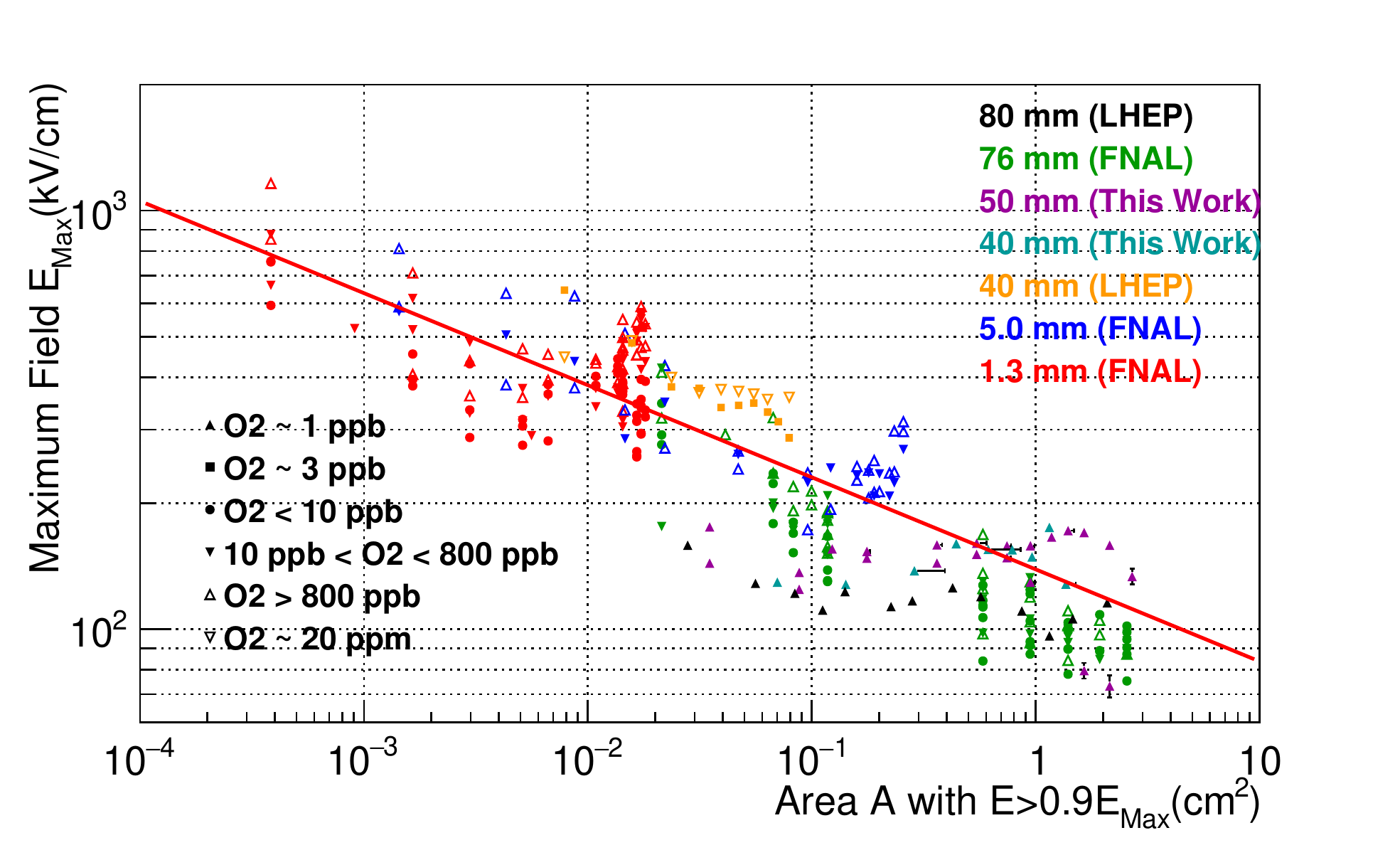}
\caption{Breakdown field versus stressed area of the cathode. The stressed area $A$ is defined as the area with an electric field intensity greater than 90\% of the maximum electric field intensity in the gap. The fit line represents the dependence $E_{max}=C\cdot A^p$ with $C = 139 \pm 5$ and $p = -0.22 \pm 0.01$. The colours correspond to different cathode sphere diameters while the marker styles correspond to different oxygen-equivalent impurity concentrations. The data are taken from \cite{FNAL_paper}~(FNAL), \cite{Blatter:2014wua}~(LHEP), and this work.}
\label{fig:powerplot}
\end{figure}

As it was shown in \cite{FNAL_paper,Gerhold94}, experimental data on breakdowns in liquefied noble gases suggest the following dependence for the maximum field at the breakdown: $E_{max}=C\cdot A^p$, where $C$ is a material-dependent constant, $A$ is the stressed area with an electric field intensity above 90\% of its maximum, and $p\approx-0.25$.
In Figure \ref{fig:powerplot}, we combine data available in literature with those obtained from this work.
Each data point is the mean value of all measurements of one run taken at the same gap distance and therefore with the same stressed area.
The global best fit gives the following values for the parameters: $C = 139 \pm 5$ and $p = -0.22 \pm 0.01$.
The statistical uncertainties represented by the error bars (smaller than the marker where not shown) are small compared to the unknown systematic uncertainties.
Indications for this are the high spread of the points around the fit line and the high reduced chi-square of 7283.

\begin{figure}[htb]
\centering
\includegraphics[width=.99\linewidth]{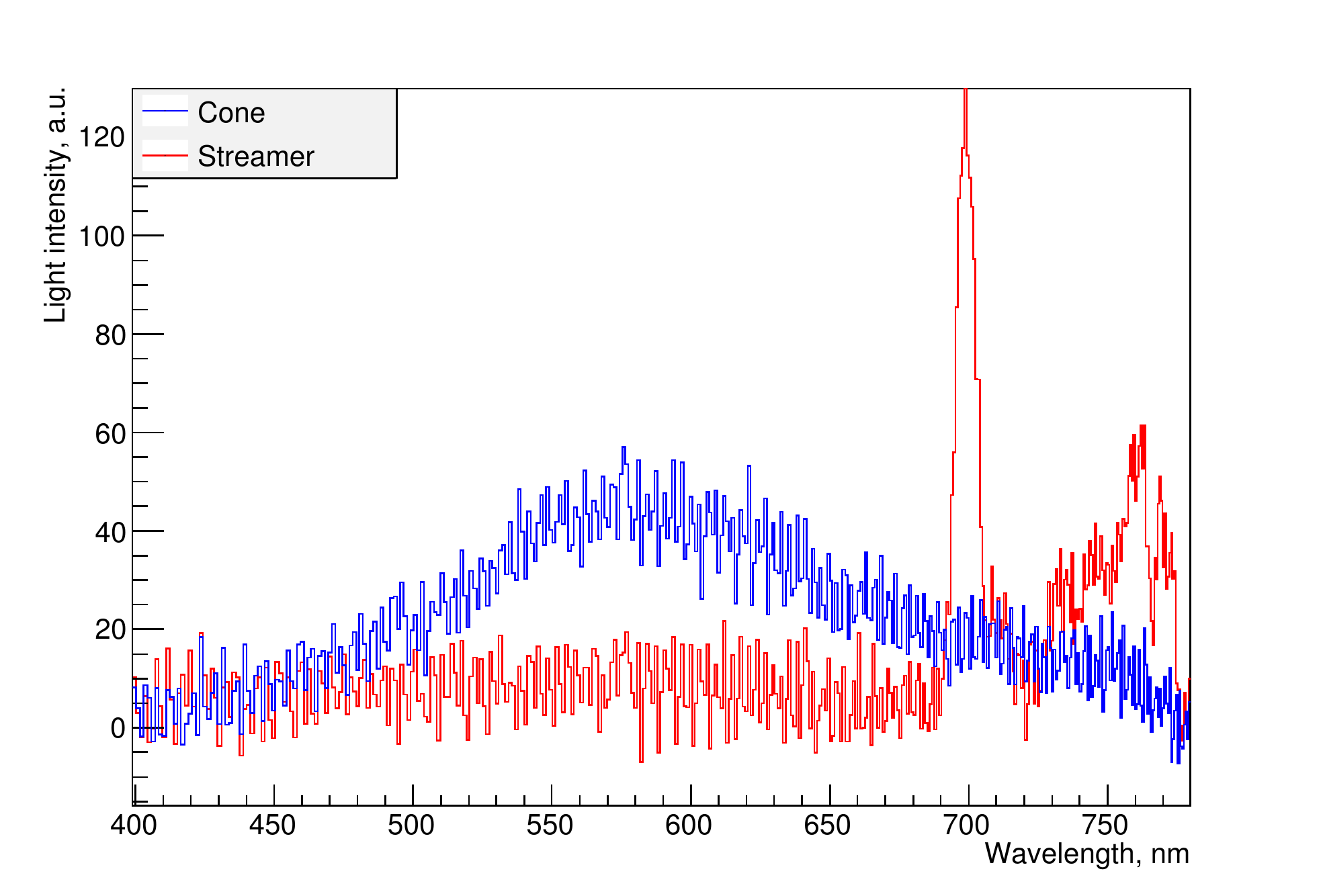}
\caption{Spectra of the field emission cone (blue) and the streamer (red) for a breakdown from a 4~cm diameter cathode at -56.2~kV and 3.0~mm apart from the anode plate. The spectra were integrated over a time of 1~ms with the spectrum of the streamer taken 2~ms after the spectrum of the cone. The blue curve is a broad continuum similar to the scintillation spectrum of liquid argon, while the red curve features a distinct peak around 700~nm which is attributed the the 3p$^5$4p--3p$^5$4s transition of neutral argon gas.}
\label{fig:spectro}
\end{figure}

Figure~\ref{fig:spectro} shows the recorded spectra of a typical event.
The spectra are integrated over 1~ms and approximately correspond to frames 3 (blue) and 8 (red) in Figure~\ref{fig:images}.
From the observation of the spectra of the emitted light and the discharge appearance, one can distinguish three phases of the breakdown development.
The first phase starts with the field emission of electrons from a point of the cathode metal surface.
The emitted electrons drift to the anode, ionizing and exciting argon atoms.
Frames 2 and 3 of Figure \ref{fig:images}, the broad current peak of the current in Figure~\ref{fig:iv}, and the blue curve in Figure~\ref{fig:spectro} show the development of the emission.
Evidence for the presence of ionization comes from the analysis of the emission spectrum in the cone formed by drifting electrons.

The emission of light by charged particles drifting in noble liquids under the influence of an electric field (electro-luminescence) gained great interest in the last years.
Recent studies in this field are well covered by~\cite{buzulutskov1, buzulutskov2} and references therein.
The red electro-luminescence, namely the peak around 700~nm of the red curve in Figure~\ref{fig:spectro}, produced by electrons drifting in argon gas is attributed to the 3p$^5$4p--3p$^5$4s transition of neutral argon~\cite{Boffard}.
The energy needed for the excitation of the electrons to the 3p$^5$4p states from the ground state in argon gas is 12.9~eV to 13.5~eV.
The ionization potential of liquid argon is 13.84~eV~\cite{Badhrees:2010zz}.
For the condensed state, only the scintillation spectrum under ionization by high-energy charged particles has been described in literature so far~\cite{Heindl}.
The electro-luminescence spectrum we measured (blue curve in Figure~\ref{fig:spectro}) exhibits a broad continuum, similar to the scintillation spectrum.
However, the center value at about 580~nm does not correspond to any of the electron transitions of neutral, singly- or doubly-ionized argon atoms.
The nearest candidate for such an emission is the residual oxygen with its strong 557.7~nm emission line.
However, if attributable to oxygen, this line has to also be observed at the later stages of the discharge, which does not take place in our measurements.

The broad width of the spectrum could be explained by smearing the energy levels into bands due to inter-atomic interactions in liquid and by the formation of exciton clusters~\cite{Bernstorff,Foerstel}.
If the energy band structure of excitons in liquid argon is continuous, as suggested by the scintillation spectrum, there might be a significant overlap of the band corresponding to the 3p$^5$4p atomic levels and the conduction band above 13.84~eV.
The presence of an observable emission at about 580~nm, in this case, is inevitably linked to a presence of ionized states.

\begin{figure}[htb]
\centering
\includegraphics[width=.99\linewidth]{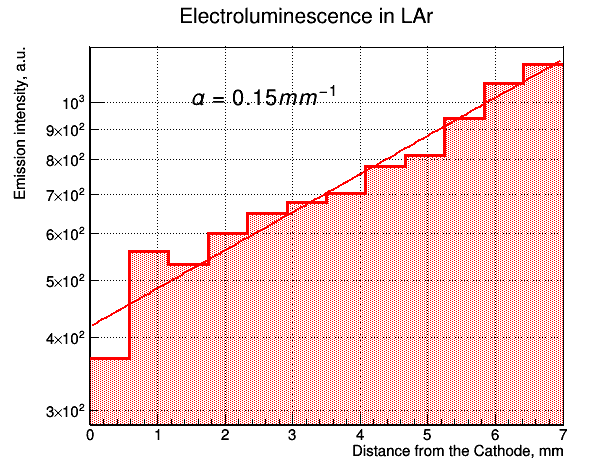}
\caption{Increasing brightness of the electro-luminescence cone as it develops towards the anode. The line represents the exponent with a fit avalanche multiplication parameter of $\alpha = (0.15 \pm 0.03)$~mm$^{-1}$ for gap conditions: voltage across the 7.0~mm gap $V = 54.0$~kV, cathode sphere diameter of 4~cm, maximum field in the gap $E_{\m{max}} = 96.1$~kV/cm, mean field in the gap $<E> = 87.0$~kV/cm.}
\label{fig:conexpo}
\end{figure}

Another signature of avalanche ionization in this phase of the breakdown is the increase of the cone brightness as it develops from the cathode towards the anode.
Figure \ref{fig:conexpo} shows this increase together with the fit avalanche multiplication parameter $\alpha = (0.15 \pm 0.03)$~mm$^{-1}$ for the following gap conditions: voltage across the 7.0~mm gap $V = 54.0$~kV, cathode sphere diameter of 4~cm, maximum field in the gap $E_{\m{max}} = 96.1$~kV/cm, mean field in the gap $<E> = 87.0$~kV/cm.
To calculate the emission intensity we summed up the raw values of all pixels of the camera image that were in a row perpendicular to the cone direction.
The distance from the cathode can be derived from the known gap distance.
From the fit we obtained the given statistical uncertainty of $\alpha$.
As only one measurement was taken, it is not possible to state anything about unknown systematic uncertainties (for instance the calibration of the camera).

As we already suggested in~\cite{Blatter:2014wua}, positive ions produced in this process drift towards the cathode, raising the surface field and provoking a rapid increase of the field emission current to values of the order of several mA.
Ions bombarding the cathode surface raise the local temperature and, after 1-2~ms, the liquid near the initial discharge point transitions to a gas phase, forming a bubble.
Both the first and the second avalanche multiplication coefficients are a few orders of magnitude higher in gas than in liquid.
Therefore, the ionization density in the gas bubble quickly rises along with the conductivity of the formed plasma.
This leads to a decrease of the electric field in the close vicinity of the field emission point and to the suppression of a further growth of the field emission current.
Accelerated electrons of the gas plasma hit the gas-liquid interface, forcing the bubble to elongate and grow.
In the region behind the head of the streamer the filament is collapsed to a diameter below 200~$\mu$m (the spatial resolution of our camera) by surface tension and electro-striction forces.
This second phase of the discharge is characterized by the growth of the streamer-like filament in liquid.
In Figure~\ref{fig:images} (frames 4 to 10), one can see the development of such a filament.
Unlike the electrons in the first phase the filament does not follow the electrostatic field lines but it rather rambles around their direction, being subject to thermodynamic fluctuations at the tip of the growing streamer where the liquid-gas transition happens.
The spectrum of the light emission from the streamer has a distinct line at about 700~nm, a characteristic feature of the plasma in argon gas.

\begin{figure}[htb]
\centering
\includegraphics[width=0.99\linewidth]{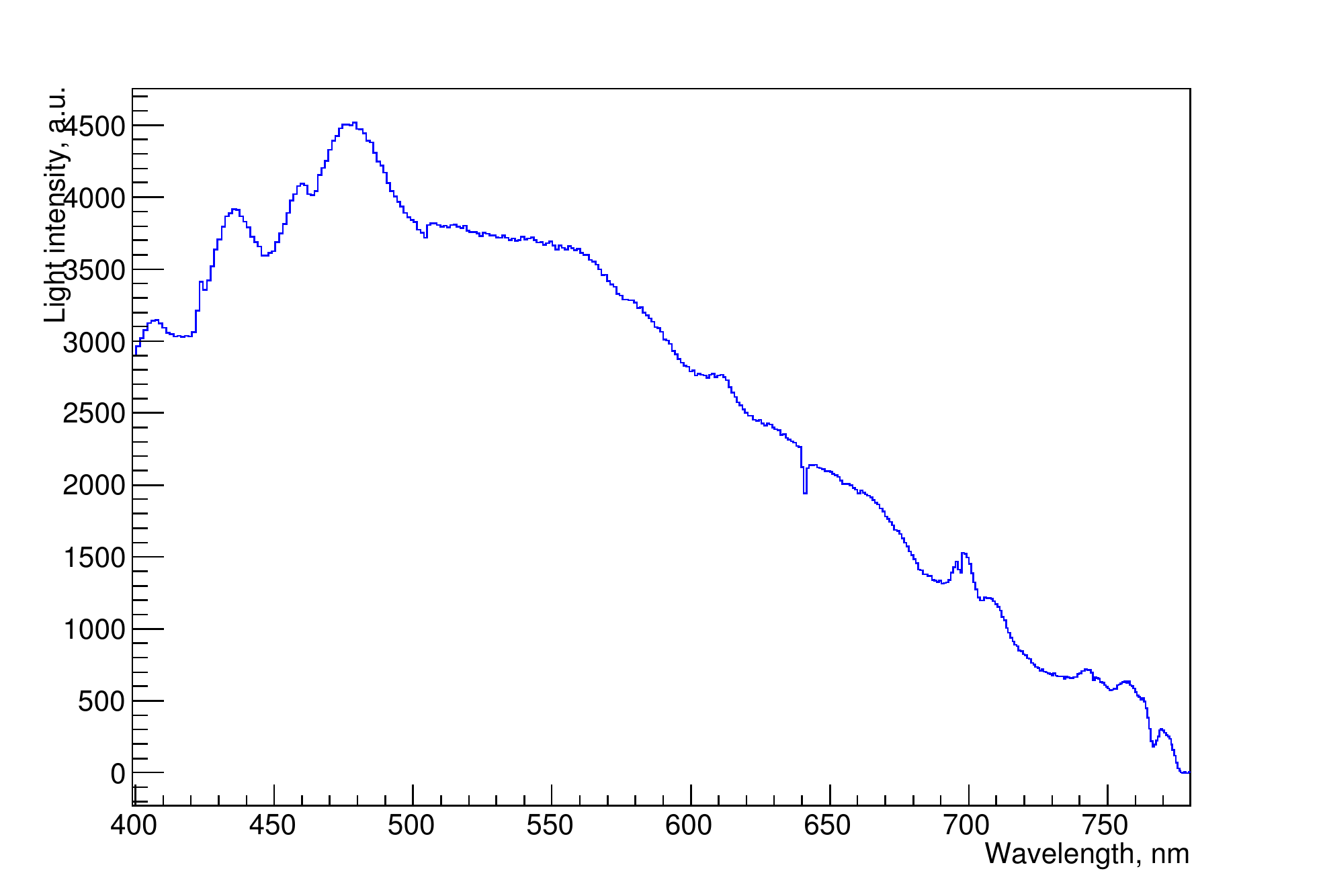}
\caption{Spectrum of the spark for a breakdown from a 4~cm diameter cathode at -39.7~kV and 4.0~mm apart from the anode plate. The spectrum was integrated over a time of 1~ms.}
\label{fig:spectro_spark}
\end{figure}

Finally, when the streamer reaches the anode a short peak of light emission is registered (frame 11 in Figure~\ref{fig:images}) with the blue-green spectral component dominating (Figure~\ref{fig:spectro_spark}).
This phase is characterized by an acoustic shock and a massive production of gas bubbles in the region of the discharge.
These effects are typical for an arc discharge in argon gas.
The spectrum of the light emission in this phase  is shown in Figure~\ref{fig:spectro_spark}. 

As it was demonstrated in~\cite{Heindl}, the transition from the liquid phase to the gas phase for scintillation manifests itself by the appearance of sharp spectral lines while in liquid, the emission spectrum is continuous and without features.
This behavior is also suggested by the two spectra in Figure~\ref{fig:spectro}.
While the spectrum is continuous during the field emission phase, there is a distinct peak at around 700~nm several milliseconds later.

It is worth mentioning that not every streamer results in a third phase spark.
For those streamers started from the side of the cathode sphere, the charge needed for streamer growth might exceed the total charge available in the system.
Such streamers extinguish before reaching the anode without an acoustic shock or any other additional effects.
On the other hand, in some cases the filament quickly transits to a third stage before it reaches the anode.
One possible explanation for this is that, if the filament current exceeds a given threshold, the filament loses its thermodynamic stability and expands into a gas bubble in which the arc discharge quickly develops afterwards.

\begin{figure}[p]
\centering
\includegraphics[height=0.43\textheight]{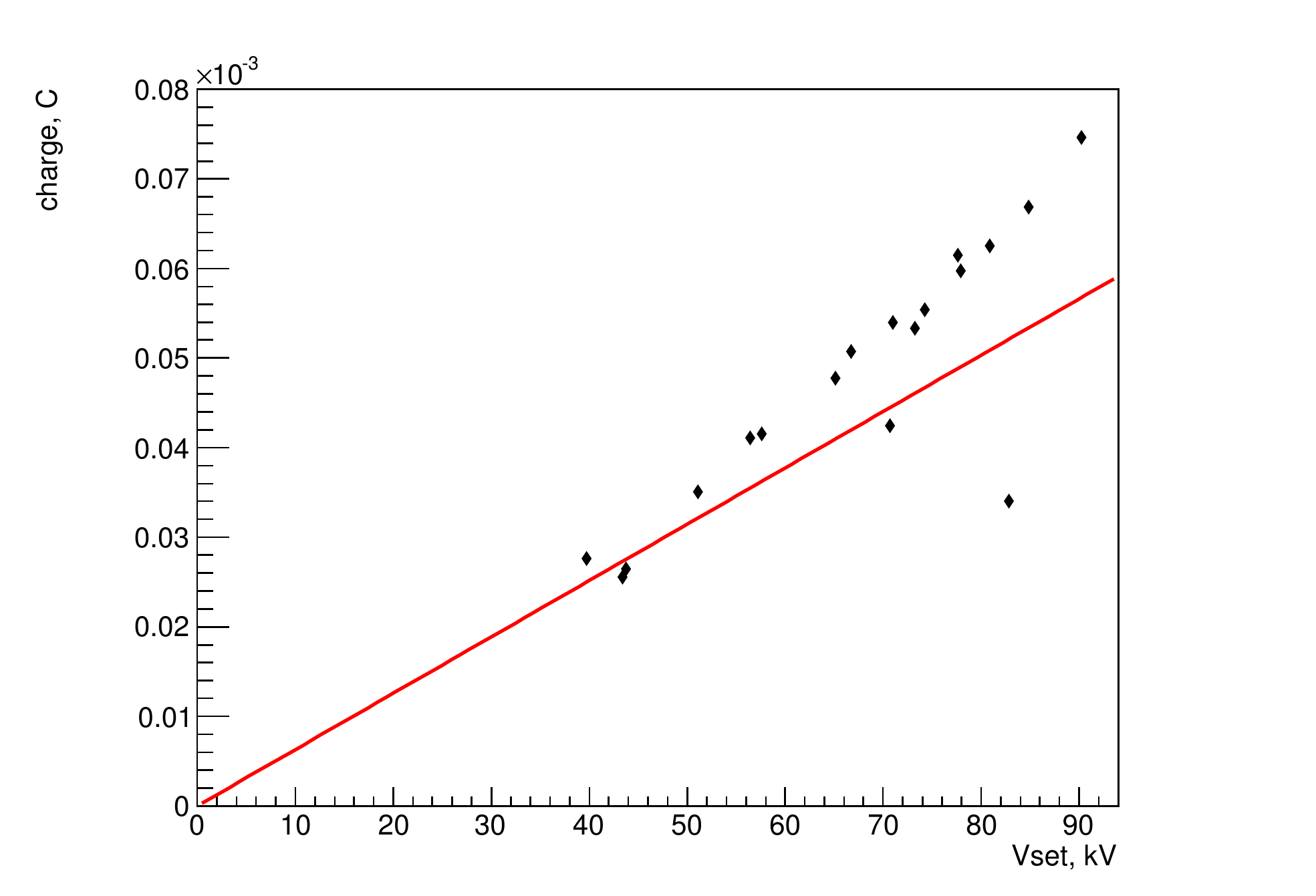}
\includegraphics[height=0.43\textheight]{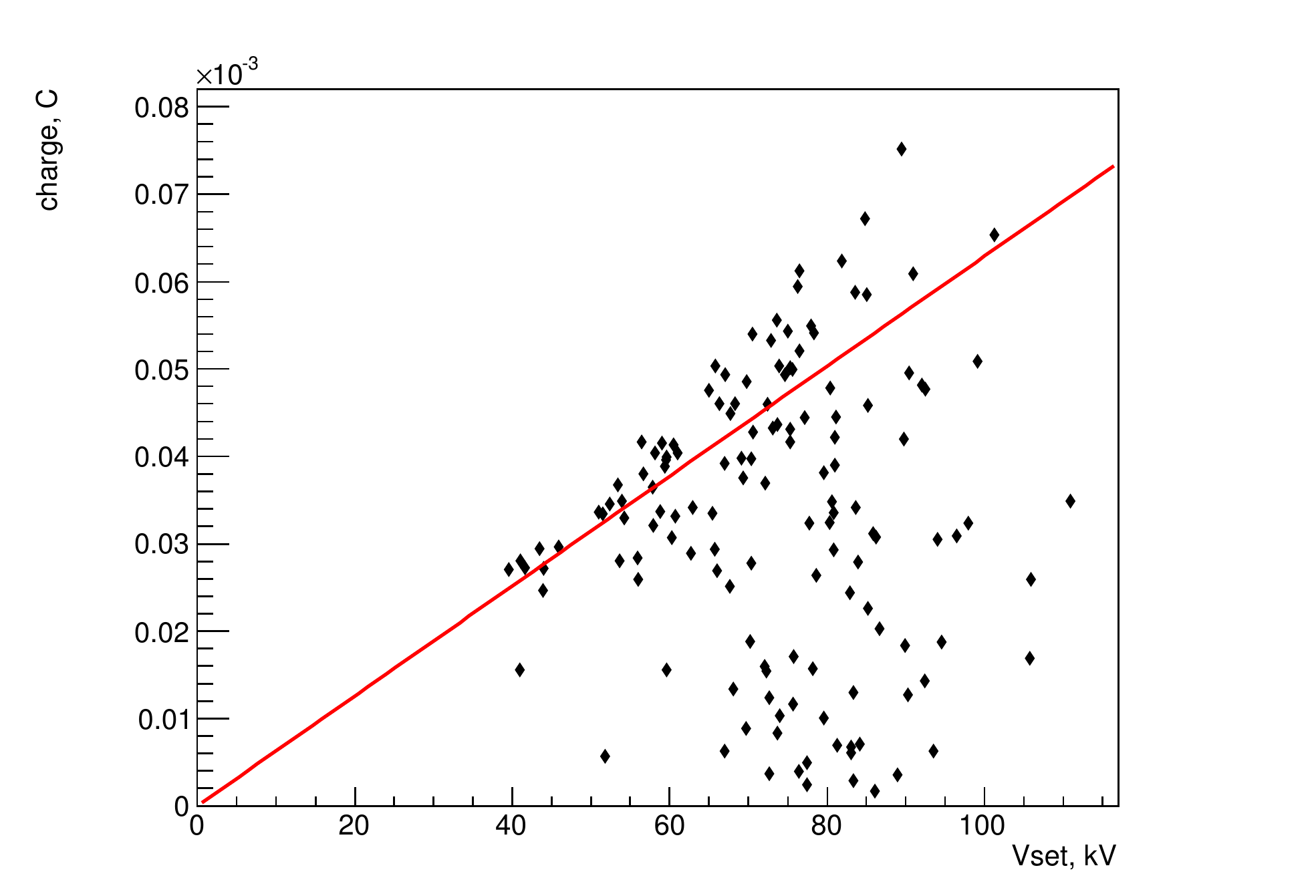}
\caption{Correlations between integrated charge and breakdown voltage \emph{Vset} for the selected events with distinguishable slow streamer phase (top) and for all events with recorded current characteristics (bottom). The red line represents the charge stored in the gap capacitance using the tuned value of the latter.}
\label{fig:chargeVsVset}
\end{figure}

\begin{figure}[p]
\centering
\includegraphics[height=0.43\textheight]{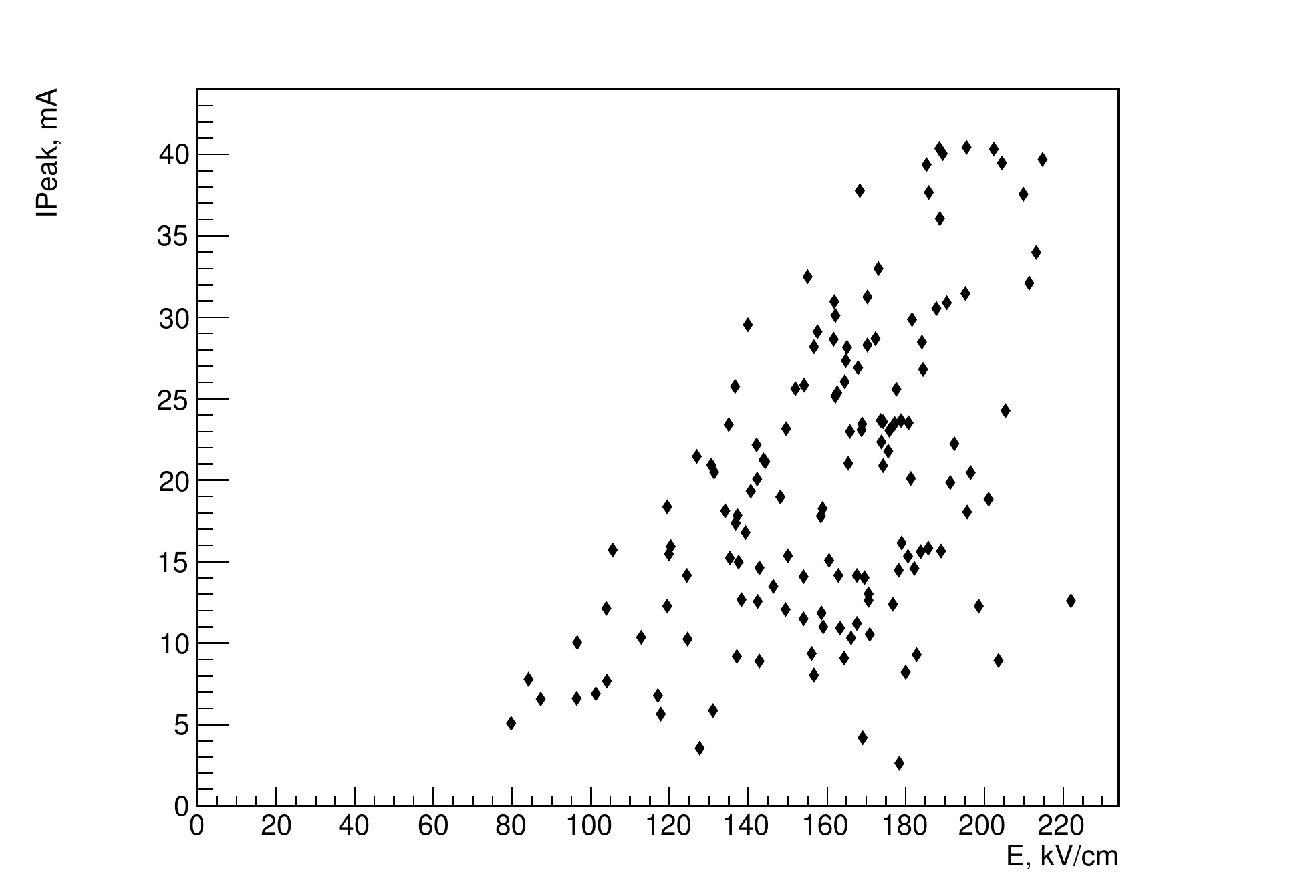}
\caption{Correlations between peak current \emph{IPeak} and maximum breakdown field \emph{E} for all events with recorded current characteristics.}
\label{fig:IPeakVsE}
\end{figure}

\begin{figure}[p]
\centering
\includegraphics[height=.43\textheight]{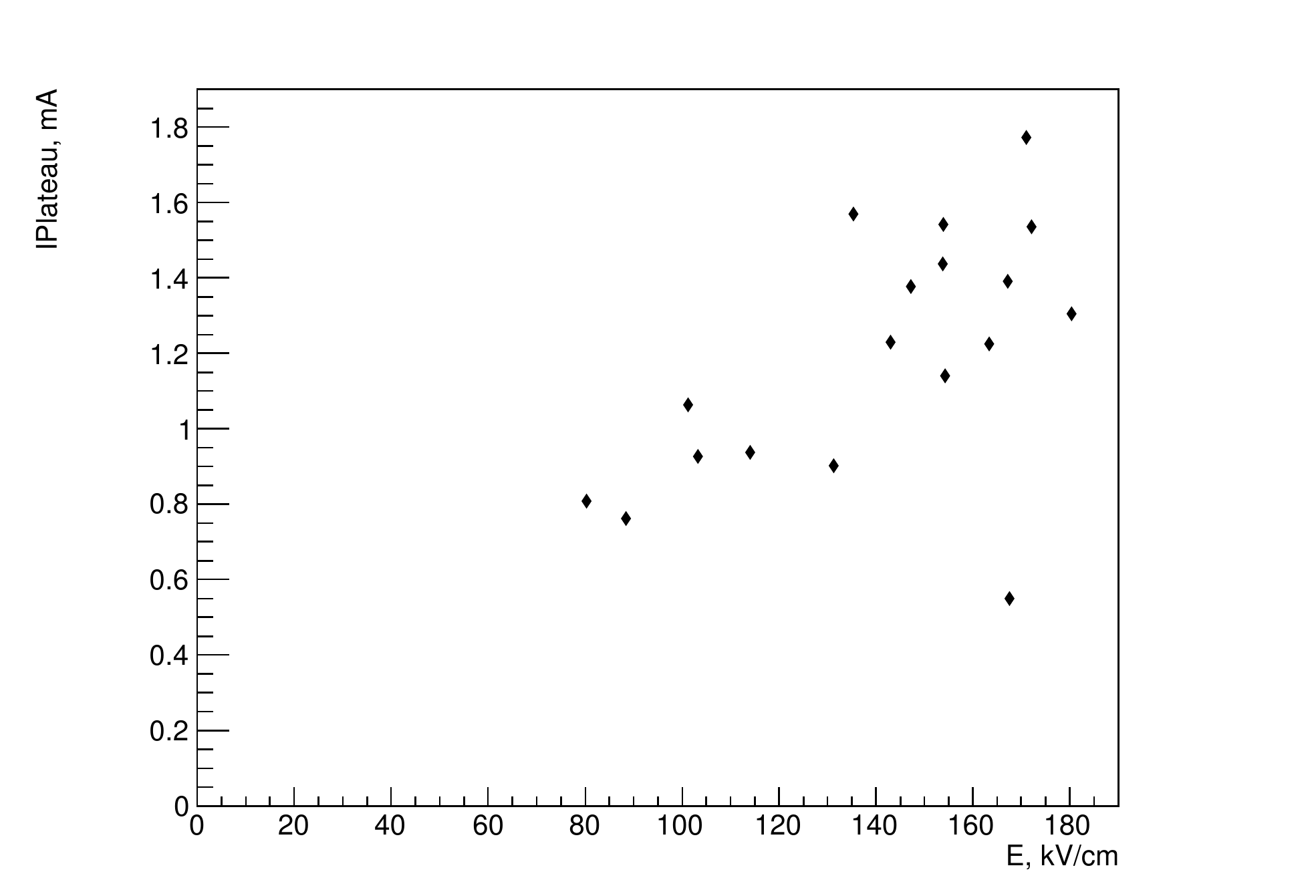}
\caption{Correlations between plateau current \emph{IPlateau} and maximum breakdown field \emph{E} for the selected events.}
\label{fig:IPlateauVsE}
\end{figure}

\begin{figure}[p]
\centering
\includegraphics[height=.43\textheight]{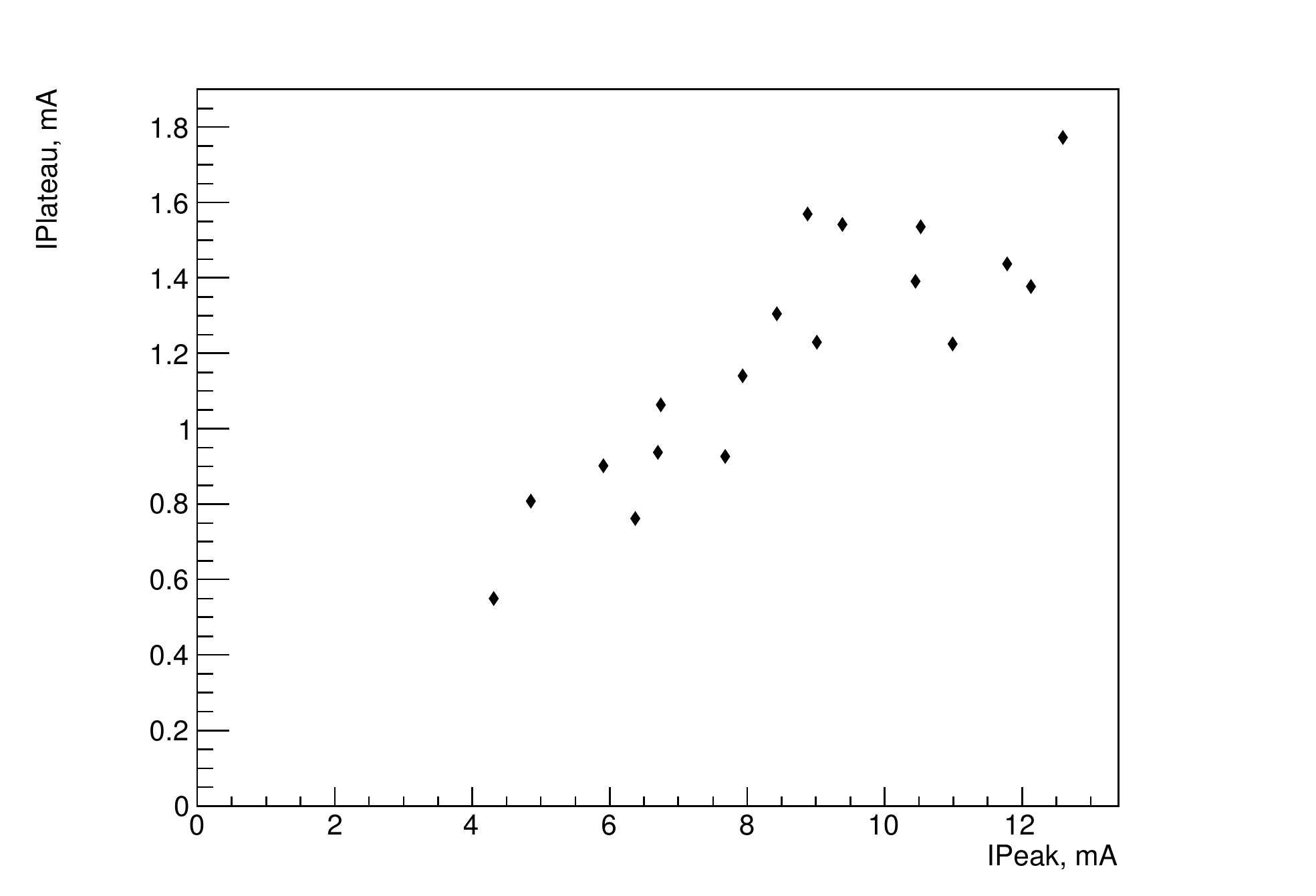}
\caption{Correlations between plateau current \emph{IPlateau} and peak current \emph{IPeak} for the selected events.}
\label{fig:IPlateauVsIPeak}
\end{figure}

\begin{figure}[p]
\centering
\includegraphics[height=.43\textheight]{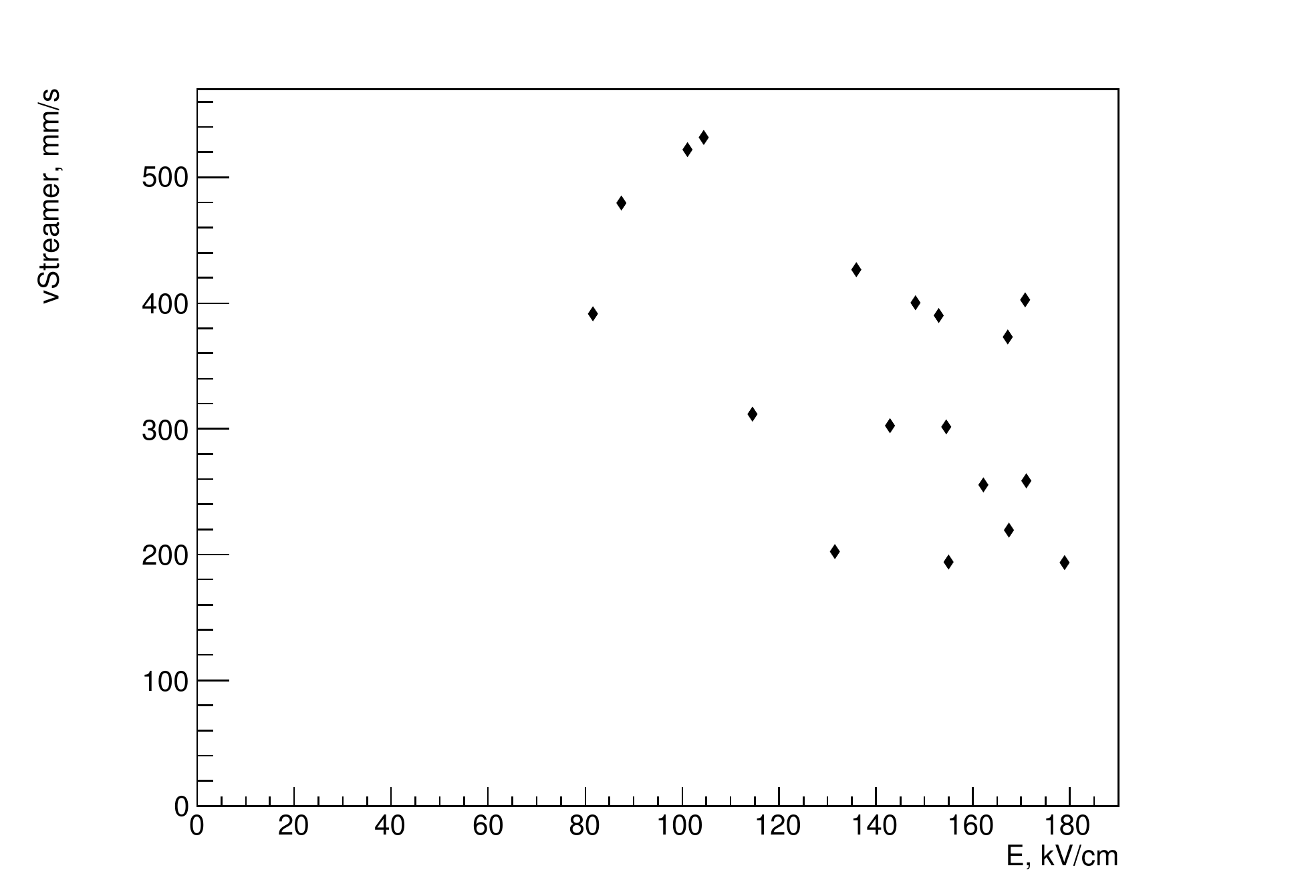}
\caption{Correlations between minimum streamer velocity \emph{vStreamer} and maximum breakdown field \emph{E} for the selected events with a distinguishable slow streamer phase.}
\label{fig:vStreamerVsE}
\end{figure}

In Figures~\ref{fig:chargeVsVset} to~\ref{fig:vStreamerVsE}, we show several correlations of measured and calculated parameters of the breakdowns.
For some of these plots, 18 events were selected with recorded current characteristics similar to Figure~\ref{fig:iv}.
As a comparison, the bottom plot of Figure~\ref{fig:chargeVsVset} shows the data of all events with current characteristics including events not possessing a distinct plateau as the one visible in Figure~\ref{fig:iv}.
The discrepancy to the total number of events in Table~\ref{table2} arises, on the one hand, because we installed the shunt resistor only in the last run and, on the other hand, since the latter was damaged after the events shown in the bottom plot of Figure~\ref{fig:chargeVsVset}.
The low number of events in the selection is due to the fact that an automated analysis of the current characteristics can only detect very long streamers.
This also explains the behavior of the charge in Figure~\ref{fig:chargeVsVset}.
The selected streamers last for several milliseconds, almost always consume the whole charge in the system and then cease without transitioning to a spark.
The slight excess in charge compared to the charge in the gap capacitance (red line) is likely supplied by the PSU before tripping.
Contrary to this, the bottom plot showing all the events contains many events that do not consume all the stored charge and result in a spark.
The good match between the red curve and the data points serves also as a crosscheck of the tuned capacitance.

Figure~\ref{fig:IPeakVsE} shows the behavior of the peak current versus the breakdown field, suggesting a proportionality between the two with the coefficient of about $60 \mathrm{\mu A}\cdot$cm/kV.
The field was calculated by dividing the breakdown voltage by the gap distance.
Therefore, this is a mean value along shortest path and does not directly apply to the selected events as most of them emerged from the side of the sphere.

Figures~\ref{fig:IPlateauVsE} and~\ref{fig:IPlateauVsIPeak} show the correlation of the current during the streamer phase (the plateau in Figure~\ref{fig:iv}) with the peak current and the breakdown field.
The plateau current clearly rises with both the breakdown field and the peak current.
Together with Figure~\ref{fig:IPeakVsE} this indicates that for higher fields, higher currents flow during both field emission as well as streamer phases.
As mentioned above, the plateau current could only be reliably detected for the selected events which is why these plots are not shown for all events.
\clearpage

Finally, Figure~\ref{fig:vStreamerVsE} depicts the dependence of the streamer velocity on the breakdown field.
Again, the velocity is only a lower limit as it was calculated by dividing the gap distance by the duration of the streamer, which is not correct for streamers emerging from the side of the sphere.
There are two distinct types of events.
While the selected streamers are rather slow (velocity $\approx$300~mm/s, independent of the field), the whole data set contains much faster events with the total time in the ns scale (not shown).
The knowledge of the streamer velocity can be applied in the design of protection circuits for future LAr~TPCs.
If a breakdown condition is detected during the streamer phase, the high voltage can be killed prior to a disruptive spark phase potentially damaging sensitive detector electronics.

\section{Conclusions}
In this work, we have extended our previous studies presented in~\cite{Blatter:2014wua} to cathode diameters of 4~cm, 5~cm and 8~cm. 
A study of the visible light emission was performed for electrical breakdown in liquid argon near its boiling point with cathode-anode distances ranging from 0.1~mm to 10.0~mm with a spherical cathode and a plane anode geometry.
Three discharge development phases were identified by observing the discharge appearance and the time development of visible light emission.
The dependence of several breakdown parameters on the critical field was studied, as well.
In particular, we found for the first time that the streamer propagation velocity is about 300~mm/s and independent of the field intensity.
The streamer phase is characterized by a current peak of the order of 5-15 mA depending on the breakdown field, followed by a plateau at about ten times lower current level.
These results provide benchmarks for the design of protection circuits for future LAr~TPCs.
By detecting the current peak caused by the streamer it would be possible to shut down the drift field before transition of the discharge to a disruptive high-current spark phase.

\section{Acknowledgments}
The authors are very grateful to Alexei Buzulutskov from the Budker Institute of Nuclear Physics SB RAS of Novosibirsk for the fruitful discussions on the mechanism of electro-luminescence in liquid argon.

%% bibliography

\end{document}